# Use of Wikipedia categories on information retrieval research: a brief review

Jesús Tramullas
Dept. of Library & Information Science
University of Zaragoza
Spain
tramullas@unizar.es

Piedad Garrido-Picazo
Dept. of Computer Science & Systems Engineering
University of Zaragoza
Spain
piedad@unizar.es

Ana I. Sánchez-Casabón
Dept. of Library & Information Science
University of Zaragoza
Spain
asanchez@unizar.es

## ABSTRACT

Information Retrieval (IR) application to the refinement and improvement of search expressions, is one of the major areas of Computer Science. That's the reason why several types of work are identified, depending on the intrinsic study of the categories structure, or its use as a tool for the processing and analysis of another documentary corpus different to Wikipedia. This paper revises them identifying not only the different category uses, and applications by adopting a systematic literature review approach of IR research but also analyzing how a knowledge organization system, developed collaboratively, is being used as a research tool in different approaches to information processing and retrieval. Surprisingly, the set of available works shows that in many cases research approaches applied and results obtained can be integrated into a comprehensive and inclusive concept of IR.

## I INTRODUCTION

With more than 32 millions of articles, Wikipedia is the primary resource of encyclopaedical information available, and millions of users consult it. It is known as a knowledge base, structured and labeled according to basic instructions and parameters. The review of the structure of the articles in Wikipedia, as well as the tools of organization and exploration of the encyclopedia, allow us to identify that one of its fundamental elements are the categories.

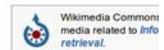

**Figure 1: Categories for "Evaluation measures (Information retrieval)" Wikipedia article (https://en.wikipedia.org/wiki/evaluation_measures_(information_retrieval).**

Wikipedia categories, an extensive collection of terms and their relations, is used by editors framing the contents within a structure for knowledge organization [1]. The categories (Fig. 1). were introduced in Wikipedia in 2003 and pages of categories and subcategories in 2004 (Fig. 2). Both elements are developed and maintained by the community collaboratively [2]. It is a system that combines a hierarchical organization with relations among different categories which creates polyhierarchies and associations. The categories structure can group items into sets and conceptual or topical subsets.

**Figure 2: Category "Information retrieval" with subcategories and pages (https://en.wikipedia.org/wiki/Category:Information_retrieval).**

This literature review aims to identify: a) the uses and applications that researchers are doing from Wikipedia category system in computer science research (RQ1); and b) to review how a knowledge organization system, developed collaboratively, is being used as a research tool in different approaches to information processing and retrieval (RQ2).

Arriving at this point, we want to point out that this proposal does not go to assess the structure, evolution nor the quality of Wikipedia categories system.

The rest of the paper sections are: section II explains the method followed. Section III the results and discussion. Finally, section IV concludes this paper with some future work challenges open.

## II METHOD

The research about the Wikipedia dynamics has resulted in the publication of papers on collaborative editing processes, behavioral patterns of user communities, vandalism, etc. [3]. The development of studies in the universe of Wikipedia has allowed the development of several systematic literature reviews, which follow different approaches [4]. Research about Wikipedia shows a variety of objectives, methods, and results. Nonetheless, it is also possible to identify in the bibliography a set of studies, which are using the textual corpus of Wikipedia and their categories structure, in research areas as information retrieval, document classification, semantics, and ontologies, or social labeling.

The methodology used to study the reference corpus has been the systematic literature review. We opted for a qualitative study, selecting specific papers to review, rather than by quantitative descriptive research or bibliometric approaches. The compilation of bibliographic data was carried out through reference query about the topic under investigation available in Web of Science and Scopus. This approach has been adopted as proposed by Okoli & Schabram [5] for the study of research on Wikipedia and has already been applied previously by the authors [6].

During the first phase, we selected the information sources and the search expression. Queries on Scopus and Web of Science (WoS) were held between November 2017 and January 2018, using the question "Wikipedia" and "categories," in the title, keyword, and abstract fields, and limited the search to papers published between 2002 and 2017. There were obtained 666 results in Scopus and 311 in Web of Science. In both cases, the first ones released were in 2006. There were not consulted, by the objectives and limits for this work, neither the digital libraries of ACM nor IEEE, by repeating content. Neither Google Scholar, due to the impossibility of limiting searches to specific positions of documents or their bibliographic references.

In a second phase, once we have obtained the raw data from the references, we processed the datasets. First, both datasets have merged, to continue with the identification and elimination of duplicate documents. This task has reduced the number of papers to 680. Subsequently, a qualitative selection of the works has been carried out, considering the thematic content identification of the different studies, by reviewing the titles, abstracts, and author keywords assigned to the same. The selection criteria established was the use or research of categories as an essential element of the work reviewed in each case. Each paper was considered by three authors independently. In case of discrepancy, the article is examined to determine their inclusion or rejection by a majority.

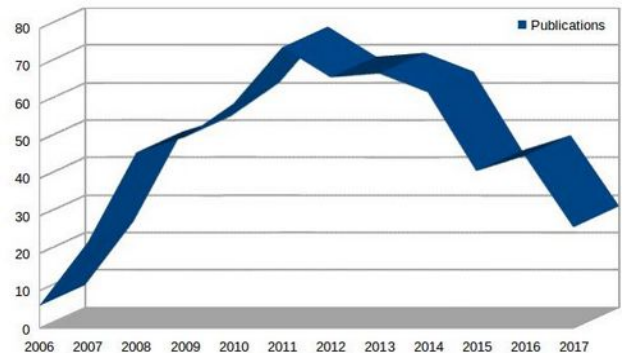

Figure 3: **Number of publications by year**

This type of selection does not allow to assess the quality nor to evaluate the quality of the contributions or their impact. The filtering carried out has made it possible to eliminate from the set those papers whose use or reference to Wikipedia categories it was not directly related to the objectives of this work. Finally, the number of selected works amounted to 546. Figure 3 shows the number of publications published by year. Studies and selected papers have been reviewed to identify in them the use made of Wikipedia categories system. We have defined the application context, the method used and the results obtained, to delineate the lines of research which have used the categories of Wikipedia as an integral part thereof. Finally, it has proceeded to the description of the results and obtain provisional conclusions. All bibliographic references collected and reviewed in the study are open access to specific groups in Mendeley and Zotero (Table 1).

Table 1: **URLs for open bibliography data**

| Mendeley | https://www.mendeley.com/community/ research-on-wikipedia-categories/ documents/ |
| --- | --- |
| Zotero | https://www.zotero.org/groups/1543457/ research_on_wikipedia_categories |

## III RESULTS AND DISCUSSION

The results obtained from the qualitative review which has been carried out demonstrate the variety of approaches, uses, and applications that researchers make with the Wikipedia categories structure. RQ1 answered in the affirmative, although some problems arise. This wealth implies a limit for the qualitative analysis proposed: the combination of techniques, approaches, and applications existing in research work makes it impossible to establish precise divisions among types of publications. While we may find papers ascribed to a subject (for example, generation of ontologies), works which also generate them also combined with other information retrieval techniques. These investigations can take place in a generic context, or specific domains. Accordingly, a qualitative treatment as it addressed in this work should be limited to delineate identified topics and research areas. To investigate the relationships between issues and possible divisions or types including, it would be necessary to apply techniques based on numerical analysis of the bibliographic corpus as developed by Smiraglia and Cai [7]).

In response to the indicated limitation, the qualitative review allows establishing the first division. Firstly, studies that analyzed the category system itself within the context of Wikipedia (covering aspects such as organization and lifecycle of content, structure, user community perspective, or the evolution and improvement of the category system). Secondly, those papers that use Wikipedia categories in the context of studies on different aspects of information processing, usually on documentary corpus independent of Wikipedia (collections of documents of different types and thematic, web pages, messages on social networks, etc.). About 80-90% of the reviewed studies belong to this second group. It is necessary to emphasize the presence of studies that use an ad-hoc corpus, as well have been generated or extracted from the Wikipedia itself. All the studies reviewed could be framed in any of the four research categories on Wikipedia listed by Nielsen [8]. Within this second group, we identified the following major research areas or topics:

• Information Retrieval: those proposals which use categories in different processes and techniques of IR, both as regards the formulation of search expression, its refinement, and improvement, or to the filtering of query results. Should be highlighted both its use in performance evaluation as well as recommendation processes.

• Entity processing: particular interest, in quantity, awaken work seeking to identify entities (named entities) in documents. The application of these studies is extensive since they serve to determine semantic relationships among terms, solve disambiguation, or integrate classifications and taxonomies. They may appear related to research on natural language processing and even the development of dictionaries.

• Indexing and classification of document corpus: categories or specific subsets are tools to proceed with the indexing and ranking of document sets, usually within particular contexts or domains. A subset of these works is made up of those who use the categories to label textual documents, within the framework of automatic indexing processes. Perhaps this may include jobs that produce specialized corpus automatically, using the Wikipedia categories in turn.

• Creating and using taxonomies: it is one of the most classic uses. Wikipedia categories are extracted from their context to form new schemes, applied in specific domains or combined with the use of other taxonomies. About this approach, we have identified some papers that propose creating classifications of classical structure, as hierarchical classifications or thesauri. On numerous occasions, taxonomies created are integrated into processes corpus indexing and rating and collected in the previous point.

• Creating and using ontologies: the second of the traditional uses. About 15% of the reviewed works deal with the creation and use of taxonomies and ontologies from the Wikipedia category system. As taxonomies indicated in the previous point, they are used in document classification processes, but also in ontological engineering and development of semantic relationships between entities.

• Semantic treatment: this group included different approaches which characterize the principles and techniques of the semantic web, methods such as the creation of graphs of categories, creating trees and schemata, extracting triplets, the identification of meaningful relationships among terms and its semantic use, etc. Ontologies, although an integral part of the semantic web, have been included in a separate group because of their importance.

## IV CONCLUSIONS AND FUTURE WORK

Wikipedia is having a significant influence in the way that users approach the resolution of their information problems and needs [9]. Scientific research does not oblivious to the importance of this information resource, which finds in Wikipedia a high-value test bank [10]. The results obtained in our study are very similar to those offered by [11]. Starting from the premise of Mehdi et al. analyze a smaller sample, and our research topics proposal is different. If computer science is the academic discipline that publishes the most works about Wikipedia, it is necessary to emphasize that also journals and conferences are those that more citations received in Wikipedia [12].

The first conclusion drew from the study is that Wikipedia is a hot topic for different research fields, both in its internal aspects and external use of its data in other areas and research approaches. Wikipedia is an essential field of research for different areas of computer science, in general, and information retrieval, in particular. Detected significant topics offer a close relationship between them, reflecting the significant outstanding issues on information retrieval. Secondly, it was necessary to emphasize its use as a tool of support and validation in different types of approaches to the study and analysis of documentary corpus, including studies about information processing, classification, and retrieval. When dealing with Wikipedia articles and its documentary corpus category system in continuous evolution, we note that results obtained in different studies may change, in the medium or long term, by the development of external and internal factors to the encyclopedia itself.

The work developed also makes it possible to detect problems that affect the method used. The variety of terms used by researchers highlights an underlying issue to systematic reviews, as is the disparity of opinion of the authors in the drafting of titles, abstracts and selecting keywords. Even in some cases, we detect the use of synonyms of terms or expressions that are not. Regarding the classification and identification of the content of the works with greater precision, shows that researchers resort to mixed approaches and combine methods and techniques, hindering a traditional plan, and it requires methods of automatic processing of information for best results. Now, we can detail some future work to do for the obtained data. First, to carry on and survey the results of applying text classification techniques to the corpus data, to compare with our proposal. Second, to complete the review with a quantitative or bibliometric analysis and finally, to study the research focused in applications of Computer Science to other fields such as e-health, geoinformatics, etc.

Finally, it is possible to emphasize the potential that offers the Wikipedia categories structure, as it is a universal classification scheme developed collaboratively, what contrasts with functional classification schemes developed in closed contexts. It provides a broad field both for the classification schemas validation, as for creating new ones from a perspective that allows combining both approaches to the organization of knowledge, as well as the advantages and advances of the Wikipedia categories system versus traditional classifications.